\documentclass[12pt]{article}
\usepackage{amsmath}
\usepackage{graphicx}
\usepackage{psfrag}
\usepackage{epsf}
\usepackage{enumerate}
\usepackage{natbib}
\usepackage{url} 

\usepackage{amsthm}
\usepackage{amssymb}
\usepackage{amsfonts}
\usepackage{dsfont}
\usepackage[labelsep=period]{caption}
\usepackage[OT1]{fontenc}
\usepackage[utf8]{inputenc}
\usepackage[colorlinks,citecolor=blue,colorlinks=true,urlcolor=blue]{hyperref}
\usepackage{placeins}
\usepackage{color}
\usepackage{stmaryrd}
\usepackage{mathtools}
\usepackage{enumitem}
\usepackage{footmisc}
\usepackage{resizegather}
\usepackage{nicefrac}
\usepackage[font=small,textfont=it,labelfont=bf]{caption}

\usepackage{geometry}
\usepackage{stackengine}
\stackMath
\newcommand\xxrightarrow[2][]{\mathrel{%
  \setbox2=\hbox{\stackon{\scriptstyle#1}{\scriptstyle#2}}%
  \stackunder[0pt]{%
    \xrightarrow{\makebox[\dimexpr\wd2\relax]{$\scriptstyle#2$}}%
  }{%
   \scriptstyle#1\,%
  }%
}}
\DeclareMathAlphabet{\mathpzc}{OT1}{pzc}{m}{it}

\theoremstyle{plain}
\newtheorem{theorem}{Theorem}

\theoremstyle{definition}

\usepackage{xr-hyper}
\usepackage{hyperref}
\newcommand{\blind}{0}

\newcommand{\supphscoreiidNormalrobust}{ S2 }
\newcommand{\suppconsistencyTimeSeries}{ S3 }
\newcommand{\suppdensityplots}{ S4 }

\newcommand{\suppdiscretehscore}{ S5 }
\newcommand{\suppidentities}{ S6 }

\newcommand{\suppconsistency}{ S7 }

\newcommand{\suppheuristicnested}{ S7.4 }
\newcommand{\suppnumericalillustrationnested}{ S7.5 }

\begin{document}

	\def\spacingset#1{\renewcommand{\baselinestretch}%
		{#1}\small\normalsize} \spacingset{1}

	
	\if0\blind
	{
		\title{\bf Bayesian model comparison with the Hyv{\"a}rinen score: computation and consistency}
		\author{Stephane Shao\thanks{\,Stephane Shao thanks the participants of Greek Stochastics iota (July 2017) for their helpful feedback. Pierre E. Jacob gratefully acknowledges support by the National Science Foundation through grant DMS-1712872 and by the Harvard Data Science Initiative. This work was also supported in part by DARPA Grant No.\! W911NF1810134. The authors also thank Christian P. Robert and Philip Dawid for their supportive and valuable comments.} , Pierre E. Jacob\footnotemark[1] , Jie Ding\footnotemark[2] , Vahid Tarokh\footnotemark[3]
			\\
			\\
			\footnotemark[1]\, Department of Statistics, Harvard University\\
			\footnotemark[2]\, School of Statistics, University of Minnesota\\
			\footnotemark[3]\, Department of Electrical and Computer Engineering, Duke University}
		\date{\today}
		\maketitle
		\vspace*{-1cm}
	} \fi

	\if1\blind
	{
		\bigskip
		\bigskip
		\bigskip
		\begin{center}
			{\LARGE\bf Bayesian model comparison with the Hyv{\"a}rinen score: computation and consistency}
		\end{center}
		\medskip
	} \fi
	
	\bigskip
	\begin{abstract}
		The Bayes factor is a widely used criterion in model comparison
		and its logarithm is a difference of out-of-sample predictive scores under the
		logarithmic scoring rule. However, when some of the candidate models
		involve vague priors on their parameters, the log-Bayes factor
		features an arbitrary additive constant that hinders its
		interpretation.  As an alternative, we consider model comparison using the Hyv{\"a}rinen score. We propose a method to
		consistently estimate this score for parametric models, using sequential
		Monte Carlo methods. We show that this score can be estimated for
		models with tractable likelihoods as well as nonlinear
		non-Gaussian state-space models with intractable likelihoods.  We prove the asymptotic consistency
		of this new model selection criterion under strong regularity assumptions in the case of non-nested models, 
		and we provide qualitative insights for the nested case. We also use existing characterizations of proper scoring rules on discrete spaces to extend the Hyv{\"a}rinen score to discrete observations. Our numerical illustrations include L\'{e}vy-driven stochastic volatility
		models and diffusion
		models for population dynamics.
	\end{abstract}
	
	\noindent%
	\vfill
	
	\newpage
	\spacingset{1.45} 
	\section{Introduction \label{sec:intro}}
	
	\subsection{Bayesian model comparison}
	Bayesian model comparison is challenging in situations where the candidate models
	involve either vague or improper prior distributions on some of their
	parameters. The Bayes factor \citep{jeffrey1939} between two models --- defined as the
	ratio of their marginal likelihoods ---  is a widely used approach to model comparison.
	If one of the candidate models includes the data-generating process, that model is termed well-specified or correct,
	and the Bayes factor can be interpreted as a ratio of odds, which updates the relative probabilities of the models being correct.
	In the misspecified or M-open setting \citep[][]{bernardo:smith:2000}, the marginal log-likelihood can be interpreted
	as a measure of out-of-sample predictive performance assessed with the logarithmic scoring rule
	\citep[e.g.][]{kassraftery1995,key1999,bernardo:smith:2000}. Scoring rules are loss functions
	for the task of predicting an observation $y$ with a probability distribution $p$, and the logarithmic 
	scoring rule quantifies predictive performance with $-\log p(y)$.
	Under regularity conditions, the Bayes factor leads to consistent
	model selection as the number of observations goes to infinity \citep[e.g.][]{dawid2011,lee2011consistency,walker2013bayesian,chib2016}.
	
	However, if any of the  models involves either vague or improper prior
	distributions on their parameters, the Bayes factor can take arbitrary values
	and becomes unreliable for any fixed sample size. This is problematic
	as vague priors are extensively used in practice,
	for instance when uniform distributions are specified on intervals of plausible values
	\citep[e.g.\!][see Section \ref{example:applicationsKangaroos}]{knape2012}.
	Improper priors also arise 
	from theoretical considerations, for instance
	as Jeffreys priors \citep[e.g.\! Chapter 3 of][]{robert2007bayesian}. Our paper takes the use of such priors by practitioners as a starting point, and addresses the question of model comparison in this context where one cannot rely on the Bayes factor.
	This limitation of the Bayes factor, sometimes
	referred to as Bartlett's paradox \citep{bartlett1957,kassraftery1995}, is a
	long-lasting challenge in Bayesian model comparison \citep[Chapter 7 of][]{robert2007bayesian},
	as it seems to suggest that prior specification should take into account the potential use (or misuse) of Bayes factors. Many approaches have been
	proposed to tackle this issue, either by modifying the Bayes factor 
	\citep[e.g.\!][]{ohagan1995,bergerPericchi1996,bergerPericchi1998,bergerPericchi2001}
	or bypassing it altogether
	\citep[e.g.\!][and references therein]{kamary:mengersen:robert:rousseau2014}.
	In this paper, we
	investigate an alternative criterion that is 1) principled for any sample size, thanks to an interpretation  in terms of predictive performance and scoring rules,
	2) enjoys asymptotic consistency properties, and 3) is robust to the arbitrary vagueness of prior distributions.
	
	Since the Bayes factor is associated with predictive performance under the logarithmic scoring rule,
	natural alternatives arise by considering other scoring rules \citep{dawid2015,dawid2016minimum}.
	We consider the \emph{Hyv{\"a}rinen score} \citep{hyvarinen2005}, 
	which is \textit{proper}, \textit{local}, and \textit{homogeneous}
	\citep{dawid2005,parry2012,ehm2012}. 
	Given $T$ observations $y_{1:T} = (y_1,...,y_T)\in\mathbb{Y}^T$ and a finite
	set $\mathcal{M}$ of candidate models, each inducing a joint marginal density
	of $(Y_1,...,Y_T)$ denoted by $p_M$ for $M\in\mathcal{M}$, we can regard the
	log-Bayes factor as a comparison of predictive sequential \citep[or \textit{prequential},][]{dawid1984} log-score $-\log p_M(y_{1:T}) = \sum_{t=1}^{T}-\log
	p_M(y_t|y_{1:t-1})$, where 
	by convention $p_M(y_1|y_{1:0})$ denotes the prior predictive distribution of $Y_1$ under model $M$. 
	%
	By contrast, for any $d_y$-dimensional observation $y\in\mathbb{R}^{d_y}$ and twice differentiable density $p$ on $\mathbb{R}^{d_y}$, the Hyv{\"a}rinen score is defined as
	\begin{align}
	\mathcal{H}(y,p)= 2\Delta_y \log p(y) + \|\nabla_y \log p(y)\|^2,
	\label{eq:definitionHscore}
	\end{align}
	where $\nabla_y$ and $\Delta_y$ respectively denote the gradient and Laplacian operators with respect to $y$. 
	We would then select the model with the smallest prequential Hyv{\"a}rinen score, defined as
	\begin{align}
	\mathcal{H}_{T}(M) = \sum_{t=1}^T \mathcal{H}\left(y_t,p_M(dy_t|y_{1:t-1})\right).
	\label{eq:definitionPrequentialHscore}
	\end{align}
	
	Homogeneity is the key property of the Hyv{\"a}rinen score which is not shared by the logarithmic scoring rule. 
	It ensures that the score does not
	depend on normalizing constants of candidate densities, hence offering robustness to vague priors and allowing for improper priors.
	For example, if $M$ denotes the toy model $Y_1,...,Y_T\,|\,\mu \stackrel{\text{i.i.d.}}{\sim} \mathcal{N}(\mu,1)$ with prior $\mu \sim \mathcal{N}(0,\sigma_0^2)$ 
	and known hyperparameter $\sigma_0>0$, then $Y_t \,|\, Y_{1:t-1} \sim \mathcal{N}\left(\mu_{t-1},\sigma_{t-1}^{2}+1\right)$ for all $t\in\{0,...,T\}$ by conjugacy,
	where
	$\sigma_t^2=(t+\sigma_0^{-2})^{-1}$ and 
	$\mu_t=\sigma_t^2 \sum_{i=1}^t Y_i$ for all $t\in\{1,...,T\}$.
	The log-score $-\log p_M(y_{1:T})$ becomes equivalent to $\log \sigma_0$ when
	$\sigma_0 \to +\infty$, and thus diverges to $+\infty$ as $\sigma_0$ increases.
	In other words, one could obtain Bayes factors that prefer virtually any
	other model over this one, by simply increasing $\sigma_0$ thus making the prior on $\mu$ arbitrarily
	vague, for any fixed number of observations $T$. 
	On the other hand, the prequential Hyv{\"a}rinen score, computed from \eqref{eq:definitionHscore} and \eqref{eq:definitionPrequentialHscore} using conjugacy, 
	converges to a finite limit as $\sigma_0\to+\infty$, so that increasing
	$\sigma_0$ can only influence the prequential Hyv{\"a}rinen score to a limited
	extent. Throughout the article, the notion of robustness to arbitrary vagueness of
	priors is to be understood in that sense. Such a robustness is desirable when models are misspecified or when the specification of vague priors is dictated by practical considerations rather than a genuine reflection of one's prior knowledge, as is sometimes the case for parameters of complex state-space models (e.g.\! see Section \ref{example:applicationsKangaroos}). The limit of $\mathcal{H}_{T}(M)$ as
	$\sigma_0 \to+\infty$ also unambiguously defines the value of the score for a flat prior $p(\mu)\propto 1$. 
	
	Without conjugacy, the calculation of the Hyv{\"a}rinen score 
	involves typically intractable integrals with respect to the sequence of partial posteriors.
	In this paper, we show how to use sequential Monte Carlo (SMC) methods
	to consistently estimate prequential Hyv{\"a}rinen scores, thereby enabling
	their use in Bayesian model comparison for general 
	parametric models. More specifically, we show that this estimation can be
	achieved for models with tractable likelihoods via SMC samplers
	\citep{chopin:2002,delmoral:doucet:jasra2006,zhou2015towards}. 
	Furthermore, the case of generic state-space models can be covered  
	by using SMC$^2$ \citep{fulop2013efficient,chopin:jacob:papaspiliopoulos2013} under the
	mild requirement that we can simulate the latent state process and evaluate the
	measurement density \citep{breto2009time,andrieu2010}, plus some integrability conditions. 
	Our second contribution is to prove that, under regularity conditions allowing
	for misspecified settings, the prequential Hyv{\"a}rinen score is consistent for model selection. Finally, motivated by an application to count-valued data
	in a population dynamics context, we propose a modified score for discrete observations 
	that builds on recent complete characterizations of proper scoring rules on discrete spaces \citep{McCarthy1956, hendrickson1971,dawid2012,dawid2017note}.
	
	This paper is organized as follows. 
	In Section
	\ref{sec:tractable_likelihoods}, we consider parametric models with tractable likelihoods. We present how the prequential
	Hyv{\"a}rinen score can be estimated via SMC samplers, and show that it leads to consistent model selection, under regularity assumptions.
	In Section \ref{sec:hscore_SSM}, we generalize the approach to nonlinear non-Gaussian state-space models, using SMC$^2$, and we present a simulation study with L\'{e}vy-driven stochastic
	volatility models.
	In Section \ref{sec:discrete}, we extend the proposed criterion to discrete observations and compare diffusion models for population dynamics. Possible limitations and directions for future research are outlined 
	in Section \ref{sec:discussion}. 
	Proofs, implementation details, and additional simulations are provided in the supplement.
	\if0\blind
	{The R code producing the figures is available at \href{https://github.com/pierrejacob/bayeshscore}{github.com/pierrejacob/bayeshscore}.
	}
	\fi

	\subsection{Terminology and notation}
	We will abbreviate the prequential Hyv{\"a}rinen score to \emph{H-score}.
	Given two models $M_1$ and
	$M_2$, the difference of their H-scores $\mathcal{H}_T(M_2)-\mathcal{H}_T(M_1)$
	will be termed the \emph{H-factor of $M_1$ against $M_2$}. 
	We define
	$\mathbb{N}^*=\mathbb{N}\setminus\{0\}$ and use the colon notation for tuples of
	objects, e.g.\! $y_{1:t} = (y_1,...,y_t)$ for all $t\in\mathbb{N}^*$, with the
	convention $y_{1:0}= \emptyset$. Unless specified otherwise, $\|\cdot\|$ 
	denotes the Euclidean norm. Each observation $y=(y_{(1)},...,y_{(d_y)})^\top$ is a vector of dimension $d_y\in\mathbb{N}^*$ and takes
	values in $\mathbb{Y}\subseteq\mathbb{R}^{d_y}$.
	Aside from Section \ref{sec:discrete}, the observations are assumed to be continuous variables. Continuous probability distributions 
	are assumed to admit densities with respect to the Lebesgue measure. 
	We let $\mathbb{P}_\star$ (resp.\! $\mathbb{E}_\star$) denote the probability (resp.\! expectation) induced by the
	data-generating mechanism of the stochastic process $(Y_t)_{t\in\mathbb{N}^*}$. We use the abbreviation $\mathbb{P}_\star$-a.s.\! for \emph{$\mathbb{P}_\star$-almost surely}.
	Assuming its existence, we let $p_\star$ denote the
	probability density or mass function associated with $\mathbb{P}_\star$. 
	When dealing concurrently with several models from a
	set $\mathcal{M}=\{M_j: j=1,...,k\}$, we use the subscript $j\in\{1,...,k\}$ to condition on a
	particular model. Each candidate model $M_j$ is parametrized by a parameter
	$\theta_j$ in a space $\mathbb{T}_j\subseteq\mathbb{R}^{d_{\theta_j}}$
	of dimension ${d_{\theta_j}}\in\mathbb{N}^*$. Explicit dependence on models
	is dropped from the notation whenever possible. 
	For a differentiable function $f$ on $\mathbb{Y}$, we use
	${\partial f(y_t)}/\partial {y_t}_{(k)}$ or $\left.{\partial f(y)}/\partial
	{y}_{(k)}\right|_{y=y_t}$ to denote the $k$-th partial
	derivative of $f$ evaluated at $y_t\in\mathbb{Y}$. 
	Hereafter, $\text{Gamma}(\alpha,\beta)$ distributions with shape $\alpha>0$ and rate
	$\beta>0$ have density $x\mapsto \beta^\alpha
	\Gamma(\alpha)^{-1}x^{\alpha-1}e^{-\beta x}$ for $x>0$; 
	a scaled inverse chi
	square distribution with degrees of freedom $\nu>0$ and scale $s>0$, denoted by
	$\text{Inv-}\chi^2(\nu,s^2)$, corresponds to the distribution of the inverse of
	a $\text{Gamma}(\nu/2,s^2\nu/2)$ variable, and has density $x\mapsto
	(\nu/2)^{\nu/2}\Gamma(\nu/2)^{-1}s^\nu x^{-(\nu/2+1)}e^{-\nu s^2/(2x)}$ for
	$x>0$; 
	$\text{NB}(m,v)$, with $v > m >0$, 
	denotes a negative binomial distribution parametrized by its mean and variance,
	i.e. with probability mass function $k\mapsto \binom{k+r-1}{k}(1-p)^r p^k$ for
	$k\in\mathbb{N}$, where $p = (v-m)/m$ and $r = m^2/(v-m)$.
	
	\section{H-score for models with tractable likelihoods}
	\label{sec:tractable_likelihoods}
	We first describe how the H-score can be estimated with SMC samplers, before turning to asymptotic properties
	and numerical investigations. The H-score defined in \eqref{eq:definitionPrequentialHscore} can be rewritten as 
	{\small
		\begin{align}
		\mathcal{H}_{T}(M) = \sum_{t=1}^T \sum_{k=1}^{d_y}\left(\,2\,\frac{\partial^2 \log p(y_t|y_{1:t-1})}{\partial {y_t}_{(k)}^{2^{\vphantom{S}}}}+ \left(\frac{\partial \log p(y_t|y_{1:t-1})}{\partial {y_t}_{(k)}}\right)^{\!2\,}\right).
		\label{eq:prequentialHscoredefinitionexpanded}
		\end{align}
	}The marginal
	predictive densities appearing in \eqref{eq:prequentialHscoredefinitionexpanded} correspond to 
	integrals with respect to posterior distributions, as $p(y_t|y_{1:t-1})=\int p(y_t|\theta,y_{1:t-1})\,p(\theta|y_{1:t-1})\,d\theta$.
	
	\subsection{Computation of the H-score using SMC}
	\label{subsec:smc}
	As noted in \citet{dawid2015}, an interchange of differentiation and integration under
	appropriate regularity conditions (see Section\suppidentities of the supplement) shows that $\mathcal{H}_{T}(M)$ is equals
	\begin{align}
	\resizebox{0.94\textwidth}{!} 
	{
		$\displaystyle\sum_{t=1}^T\sum_{k=1}^{d_y} \left(2\,\mathbb{E}_t\!\left[\frac{\partial^2 \log p(y_t|y_{1:t-1},\Theta)}{\partial{y_t}_{(k)}^{2^{\vphantom{S}}}} + \left(\frac{\partial \log p(y_t|y_{1:t-1},\Theta)}{\partial {y_t}_{(k)}}\right)^{\!2\,}\right] -\left(\mathbb{E}_t\!\left[\frac{\partial \log p(y_t|y_{1:t-1},\Theta)}{\partial {y_t}_{(k)}}\right]\right)^{\!2\,}\right),$
	}
	\label{eq:hscoreWithExpectationsMultivariate}
	\end{align}
	where the conditional expectations $\mathbb{E}_t$ are taken with respect to the posterior
	distributions $\Theta\sim p(d\theta|y_{1:t})$. The terms of the sum in \eqref{eq:hscoreWithExpectationsMultivariate}
	might not be well-defined when improper posterior distributions arise from improper priors. 
	If $\tau$ denotes the first index such that the posterior $p(d\theta|y_{1:\tau})$ is proper,
	then we would redefine the H-score as
	$\sum_{t=\tau}^T \mathcal{H}\left(y_t,p(dy_t|y_{1:t-1})\right)$. 
	This issue is not specific to the H-score, and for simplicity of exposition, we will thereafter assume that posterior
	distributions are proper after assimilating one observation.
	
	In general, expectations with respect to $p(d\theta|y_{1:t})$ for all successive $t\geq 1$ can be
	consistently estimated using sequential or annealed importance sampling \citep{neal2001} and SMC samplers \citep{chopin:2002,delmoral:doucet:jasra2006}. An SMC sampler starts by sampling a set of $N_\theta$ particles
	$\theta^{(1:N_\theta)} = (\theta^{(1)},\ldots,\theta^{(N_\theta)})$ independently from an initial distribution
	$q(d\theta)$. The algorithm then assigns weights, resamples, and moves these particles in
	order to approximate $p(d\theta|y_{1:t})$ 
	for each $t\geq 1$. 
	We can move samples from a posterior distribution to the next 
	by successively targeting intermediate distributions whose densities are
	proportional to $p(\theta|y_{1:t-1}) p(y_t|y_{1:t-1},\theta)^{\gamma_{t,j}}$,
	where $0=\gamma_{t,0} < \gamma_{t,1} < ... < \gamma_{t,J_t} = 1$ with
	$J_t\in\mathbb{N}^*$. The temperatures $\gamma_{t,j}$ can be determined
	adaptively to maintain a chosen level of non-degeneracy in the
	importance weights of the particles, e.g.\! by forcing the effective sample size to stay above a
	desired threshold or by imposing a minimum number of unique particles. The resampling steps can be performed in various ways 
	\citep[see][]{douc2005comparison,murray2016parallel,gerber2017negative},
	and the move steps with any Markov chain Monte Carlo method. 
	In the numerical
	experiments below, resampling is done with the Srinivasan Sampling Process \citep[SSP,][]{gerber2017negative},
	and move steps are independent Metropolis--Hastings steps with proposals obtained as 
	mixtures of Normal distributions fitted on the current weighted particles.
	The initial distribution $q(d\theta)$ can be taken as the uniform distribution on a set \citep[e.g.][]{fearnhead2013adaptive}, 
	as the prior distribution $p(d\theta)$ when it is proper, or more generally as an approximation of the first proper posterior distribution. 
	
	Sequential estimation of the H-score can thus be achieved at
	a cost comparable to that of estimating the log-evidence. Indeed, both can be obtained
	from the same SMC runs.
	However, numerical experiments suggest that the estimator of the H-score tends to have
	a larger relative variance than the estimator of the log-evidence, for a given number of particles. 
	This can be explained informally as follows. For the evidence,
	the Monte Carlo approaches approximate expectations of the form $\mathbb{E}[p(y_t |y_{1:t-1}, \Theta)]$ with respect 
	to the posterior $p(d\theta|y_{1:t-1})$. On the other hand, the H-score involves expectations such as 
	$\mathbb{E}[\nabla_y \log p(y_t |y_{1:t-1}, \Theta)]$ with respect to $p(d\theta|y_{1:t})$. 
	When $t$ is large, the distributions $p(d\theta|y_{1:t-1})$ and $p(d\theta|y_{1:t})$ are similar, whereas the integrands 
	$\theta \mapsto p(y_t |y_{1:t-1}, \theta)$  and $\theta \mapsto \nabla_y \log
	p(y_t |y_{1:t-1}, \theta)$ are different. In some generality, the first type of integrands
	will be easier to integrate than the second one, e.g. when the former is bounded in $\theta$
	while the latter is polynomial in $\theta$, as in Normal location models (see Section
	\ref{sec:consistencyNumericalExample}).

	\subsection{Consistency of the H-score for i.i.d.\! settings}
	\label{sec:consistencyIID}
	
	Irrespective of model misspecification, the H-score can be justified for finite samples 
	since it results from assessing predictions with a scoring rule that satisfies desirable properties such as propriety, locality, and homogeneity \citep{parry2012,ehm2012}.
	Moreover, under regularity conditions, we can show that the H-score also satisfies sensible asymptotic properties: as the number of observations grows, choosing
	the model with the smallest H-score eventually leads to selecting the model
	closest to the data-generating process in a certain sense, as made precise below. 
	A general perspective on consistency of prequential scores can be found in \citet{dawid2015}.
	
	Here we consider i.i.d.\! models and assume that $(Y_t)_{t\in\mathbb{N}^*}$ is
	a sequence of i.i.d.\! observations drawn from $p_\star$. State-space models
	and more general data-generating processes will be covered in Section
	\ref{sec:consistencySSM}. For simplicity, we focus on continuous univariate ($d_y = 1$) observations.  
	Our results will only be meaningful for models that are either non-nested, or nested with at
	most one model being well-specified. The case of well-specified nested models
	is discussed at the end of this section, with more details in Section\suppheuristicnested of the supplement.
	Our consistency result rely on the expression
	\begin{align}
	\mathcal{H}_{T}(M) = \left(\sum_{t=1}^T \mathbb{E}_t\left[\vphantom{\frac{\partial \log p(y_t|y_{1:t-1},\Theta)}{\partial y_t}}\mathcal{H}\left(y_t,p(dy_t|y_{1:t-1},\Theta)\right)\right]\right) + \left(\sum_{t=1}^T \mathbb{V}_t\left[\frac{\partial \log p(y_t|y_{1:t-1},\Theta)}{\partial y_t}\right]\right),
	\label{eq:prequentialHscoreExpectationVariance}
	\end{align}
	which follows directly from rearranging the terms in
	(\ref{eq:hscoreWithExpectationsMultivariate}), where $\mathbb{E}_t$ and $\mathbb{V}_t$ respectively denote conditional expectations and variances with respect to $\Theta\sim p(d\theta|y_{1:t})$. The key insight is that, in non-nested settings, as the
	number of observations grows and the posterior distribution
	$p(d\theta|y_{1:T})$ concentrates to a point mass, the sum of the conditional
	expectations in \eqref{eq:prequentialHscoreExpectationVariance} will eventually dominate and drive the behavior of the 
	H-score, while the sum of the conditional variances acts as a
	penalty term that becomes negligible. This penalty term only becomes crucial when comparing well-specified nested models,
	as discussed at the end of this section. 
	
	The result below considers model selection consistency for two i.i.d.\! models
	$M_1$ and $M_2$, each describing the
	data respectively as $Y_1,...,Y_{T}\,|\,\theta_j \overset{\text{i.i.d.}}{\sim}
	p_j(dy|\theta_j)$, with parameter $\theta_j\in\mathbb{T}_j$ and prior density
	$p_j(\theta_j)$, for $j\in\{1,2\}$. 
	\begin{theorem}
		\label{theorem:consistencyIID_all_in_one}
		Assume $(Y_t)_{t\in\mathbb{N}^*}$ is a sequence of i.i.d.\! draws from
		$p_\star$. Assume $M_1$ and $M_2$ both satisfy the following conditions, where
		models are omitted from the notation and probabilistic
		statements are $\mathbb{P}_\star$-almost sure:
		\begin{enumerate}[label=(\alph*)]
			\itemsep0.25em 
			\item\label{cond:1a} For all $t\in\mathbb{N}^*$ and $y_{1:t}\in\mathbb{Y}^{t}$, $\theta\mapsto p(y_t|\theta) \, p(\theta|y_{1:t-1})$ is integrable on $\mathbb{T}$. 
			\item\label{cond:1b} For all $t\in\mathbb{N}^*$ and $\theta\in\mathbb{T}$, $y_t\mapsto p(y_t|\theta)$ is twice differentiable on $\mathbb{Y}$.
			\item\label{cond:1c} For all $t\in\mathbb{N}^*$, there exist integrable functions $h_{1,t}$ and $h_{2,t}$ such that, for all $(y_{1:t},\theta)\in\mathbb{Y}^{t}\times\mathbb{T}$, $\left|p(\theta|y_{1:t-1})\,\partial p(y_t|\theta) / \partial {y_t}\right|\leq h_{1,t}(\theta)$ and $\left|p(\theta|y_{1:t-1})\,\partial^2 p(y_t|\theta) / \partial {y_t}^2\right|\leq h_{2,t}(\theta)$.
			\item\label{cond:1d} There exists $\theta^\star\in\mathbb{T}$ such that, if $\Theta_t\sim p(d\theta|Y_{1:t})$ for all $t\in\mathbb{N}^*$, then $\Theta_t\xxrightarrow[t \to +\infty]{\mathcal{D}}\theta^\star$.
			\item\label{cond:1e} There exist a constant $L>0$ and a neighborhood $\,\mathcal{U}_{\theta^\star}$ of $\theta^\star$ such that, for all $t\in\mathbb{N}^*$, $\theta\mapsto\mathcal{H}\left(Y_t,p(dy_t|\theta)\right)$  and $\theta\mapsto \partial \log p(Y_t|\theta)/\partial y_t$ are $L$-Lipschitz functions.
			\item\label{cond:1f} There exist $\alpha_1>1$ and $\alpha_2>1$ such that $\,\sup_{t\in\mathbb{N}^*}\mathbb{E}\left[\,|\mathcal{H}\left(Y_t,p(dy_t|\Theta_t)\right)|^{\alpha_1}\,|\,Y_{1:t}\,\right]< +\infty$ and $\,\sup_{t\in\mathbb{N}^*}\mathbb{E}\left[\left(\partial \log p(Y_t|\Theta_t)/\partial y_t\right)^{2\,\alpha_2}\,|\,Y_{1:t}\,\right]< +\infty$, where the conditional expectations are with respect to the posterior distribution $\Theta_t\sim p(d\theta|Y_{1:t})$.
			\item\label{cond:1g} $\mathbb{E}_\star\left[\left|\mathcal{H}\left(Y,p(dy|\theta^\star)\right)\right|^{\vphantom{S}}\right] < +\infty\;$ and $\;\displaystyle p_\star(y)\,{\partial \log p(y|\theta^\star)}/{\partial y} \xxrightarrow[|y|\to +\infty]{} 0$.
		\end{enumerate} 
		We also assume that the data-generating density $p_\star$ is such that $y\mapsto p_\star(y)$ is twice differentiable and $\mathbb{E}_\star[\left|\mathcal{H}\left(Y,p_\star(dy)\right)\right|^{\vphantom{S}}] < +\infty$. If all the conditions are met, then we have
		\begin{align}
		\label{theorem:consistencyIID}
		\frac{1}{T}\left(\mathcal{H}_{T}({M_2})-\mathcal{H}_{T}({M_1})^{\vphantom{S^S}}\right) \;\xxrightarrow[T \to +\infty]{\mathbb{P}_\star-a.s.}\;D_\mathcal{H}(p_\star,M_2)-D_\mathcal{H}(p_\star,M_1)\, ,
		\end{align}
		where, for each $j\in\{1,2\}$, the quantity
		\begin{align}
		\label{eq:DivergenceDeltaDefinitionIID}
		D_\mathcal{H}(p_\star,M_j)=\mathbb{E}_\star\left[\mathcal{H}\left(Y,p_{j}(dy|\theta^\star_j)\right)^{\vphantom{S}}\right] - \mathbb{E}_\star\left[\mathcal{H}\left(Y,p_\star(dy)\right)^{\vphantom{S}}\right]
		\end{align}
		satisfies $D_\mathcal{H}(p_\star,M_j)\geq 0$, with
		$D_\mathcal{H}(p_\star,M_j)=0$ if and only if $p_{j}(y|\theta^\star_j)=p_\star(y)$ for all $y\in\mathbb{Y}$. 
	\end{theorem} 
	
	The assumptions listed in Theorem \ref{theorem:consistencyIID_all_in_one} are
	strong, which allows for more intuitive proofs. Our numerical
	experiments suggest that \eqref{theorem:consistencyIID} can hold when
	these conditions are not met (e.g.\! see Section
	\ref{sec:consistencyNumericalExample}). Conditions \ref{cond:1a} to
	\ref{cond:1c} ensure the validity of
	\eqref{eq:prequentialHscoreExpectationVariance}; \ref{cond:1d} assumes the
	concentration of the posterior to a point mass; \ref{cond:1e} to \ref{cond:1f}
	ensure suitable convergence of posterior moments; and \ref{cond:1g} ensures the
	strict propriety of the H-score and its definiteness for $p_\star$. Further
	discussion on these conditions and detailed proofs are provided in Section\suppconsistency of the supplement.  
	
	Theorem \ref{theorem:consistencyIID_all_in_one} provides insights 
	into the asymptotic behavior of the H-score.
	Using integration by parts, we have
	\begin{align}
	D_\mathcal{H}(p_\star,M_j) = \int \left(\frac{\partial \log p_\star(y)}{\partial y}-\frac{\partial \log p_j(y|\theta_j^\star)}{\partial y}\right)^2 p_\star(y)dy,
	\label{eq:DH_Fisherdivergence}
	\end{align}
	so that $D_\mathcal{H}(p_\star,M_j)$ can be interpreted as a divergence between the data-generating distribution $p_\star$ and model $M_j$. As long as $\mathbb{E}_\star\left[\mathcal{H}\left(Y,p_{1}(dy|\theta^\star_1)\right)\right]\neq
	\mathbb{E}_\star\left[\mathcal{H}\left(Y,p_{2}(dy|\theta^\star_2)\right)\right]$, the H-score asymptotically chooses the model closest to the data-generating distribution
	$p_\star$ with respect to the divergence $D_\mathcal{H}$.
	In particular, if $M_1$ is
	well-specified and $M_2$ is misspecified, then $D_\mathcal{H}(p_\star,M_1)=0 <
	D_\mathcal{H}(p_\star,M_2)$, which leads to
	$\mathcal{H}_{T}({M_2})-\mathcal{H}_{T}({M_1})>0$ for all sufficiently large $T$,
	$\mathbb{P}_\star$-almost surely. In other words, the H-score eventually chooses a well-specified model $M_1$ over a
	misspecified model $M_2$. 
	
	The divergence $D_\mathcal{H}(p_\star,M_j)$ appearing in \eqref{eq:DH_Fisherdivergence} is sometimes referred to as the \emph{relative Fisher information divergence} between $p_\star$ and $p_j(dy|\theta_j^\star)$ \citep[e.g.\!][]{walker2016bayesian,holmes2017assigning}. It should be contrasted to the divergence associated with the log-score: under similar assumptions, one can prove \citep[e.g.][]{dawid2011} that
	\begin{align*}
	\frac{1}{T}\left(\left(-\log p_2(Y_{1:T})^{\vphantom{S^s}}\right)^{\vphantom{S^s}}-\left(-\log p_1(Y_{1:T})^{\vphantom{S^s}}\right)\right) \;\xxrightarrow[T \to +\infty]{\mathbb{P}_\star-a.s.}\;\text{KL}(p_\star,M_2)-\text{KL}(p_\star,M_1),
	\end{align*}
	where $\text{KL}(p_\star,M_j) = \mathbb{E}_\star\left[-\log p_j(Y|\theta_j^\star)\right] - \mathbb{E}_\star\left[-\log p_\star(Y)\right]$ denotes the Kullback-Leibler divergence between $p_\star$ and $p_j(dy|\theta_j^\star)$. In other words, the log-score $-\log p_j(Y_{1:T})$ asymptotically favors the model that is the closest to $p_\star$ with respect to the Kullback-Leibler divergence $\text{KL}(p_\star,M_j)$, whereas the H-score $\mathcal{H}_{T}(M_j)$ asymptotically favors the model that is the closest to $p_\star$ with respect to the divergence $D_\mathcal{H}(p_\star,M_j)$. 
	
	When only one of the candidate models is well-specified, the log-Bayes factor
	and the H-factor both agree on consistently selecting it. When both $M_1$ and $M_2$ are misspecified, each criterion selects a
	model according to its associated divergence. Despite being related
	\citep[e.g.\! ][and references therein]{bobkov2014bounds}, the geometries
	induced by these divergences differ, leading the log-Bayes
	factor and the H-factor to select possibly different models (see case 3 in
	Section \ref{sec:consistencyNumericalExample}). In the presence of informative priors,
	deciding which score to use in such misspecified settings is
	then a matter of preferences and further practical considerations; in this article
	we focus on the case of vague priors for which Bayes factors are not recommendable,
	as discussed earlier.
	
	If
	$\mathbb{E}_\star\left[\mathcal{H}\left(Y,p_{1}(dy|\theta^\star_1)\right)\right]=
	\mathbb{E}_\star\left[\mathcal{H}\left(Y,p_{2}(dy|\theta^\star_2)\right)\right]$, the limit in \eqref{theorem:consistencyIID} becomes 0 and calls for a more careful look at the higher order penalty term formed by the conditional variances in 
	\eqref{eq:prequentialHscoreExpectationVariance}. Such a refinement is needed if $M_1$ is nested in $M_2$, in the sense of Eq.\! (9) in \cite{bergerPericchi1996}, and both models are well-specified. In other words, we have $\mathbb{T}_2 = \{(\theta_1,\eta)\in\Xi_1\times\Xi_2\}\subseteq\mathbb{R}^{k_1}\times\mathbb{R}^{k_2-k_1}$ and $\mathbb{T}_1 \subseteq \Xi_1$ for some $k_1,k_2\in\mathbb{N}$ with $k_2 > k_1 > 0$, and there exists $\eta_1^\star\in\Xi_2$ such that $p_1(y|\theta_1) = p_2(y|\theta_1,\eta_1^\star)$ for all $(y,\theta_1)\in\mathbb{Y}\times\mathbb{T}_1$. There also exists $\theta_1^\star\in\mathbb{T}_1$ such that $ p_\star(y) = p_1(y|\theta_1^\star) = p_2(y|\theta_2^\star) $ for all $y\in\mathbb{Y}$, where $\theta_2^\star = (\theta_1^\star,\eta_1^\star)$. The particular case of nested Normal linear models is discussed in Sections 8 and 9 of \citet[][]{dawid2015}. Under regularity conditions, and if the parameters are orthogonal such that $\mathbb{E}_\star[\nabla_{\eta}\nabla_{\theta_1}\log p_2(Y|\theta_1^\star,\eta_1^\star)] = 0$, we conjecture that
	\begin{align*}
	\mathcal{H}_{T}({M_2})-\mathcal{H}_{T}({M_1}) = \delta_{21} \log T + o(\log T),
	\end{align*}
	as $T\to\infty$, in $\mathbb{P}_\star$-probability, where the difference $\delta_{21}$ in model dimensions appears as
	\begin{align*}
	\delta_{21} = \mathbb{E}_\star\left[\left(\nabla_{\eta}\frac{\partial \log p_2(Y|\theta_2^\star)}{\partial y}\right)^\top\, \mathbb{E}_\star[-\nabla^2_{\eta}\log p_2(Y|\theta_2^\star)]^{-1}\, \left(\nabla_{\eta}\frac{\partial \log p_2(Y|\theta_2^\star)}{\partial y}\right)\right]>0.
	\end{align*}
	This would imply that $\mathcal{H}_{T}({M_2})-\mathcal{H}_{T}({M_1})\to+\infty$ as $T\to+\infty$, in $\mathbb{P}_\star$-probability, so that the H-score asymptotically chooses the model $M_1$ of smaller dimension, similarly to the log-Bayes factor for which $\log p_1(Y_{1:T})-\log p_2(Y_{1:T}) = (1/2)(k_2-k_1)\log T + o(\log T)$  under suitable assumptions \citep[e.g.][]{moreno2010,rousseau2012,chib2016}. Heuristic justification and numerical illustration of this postulate are provided in Sections\suppheuristicnested and\suppnumericalillustrationnested of the supplement. We leave more formal studies of the H-score in nested well-specified settings for future research.
	
	As an aside, we need to contrast the prequential approach described in \eqref{eq:definitionPrequentialHscore} with a
	batch approach, where one would assess the predictive performance of model $M$ at once via $\mathcal{H}_{T}^{batch}(M) = \mathcal{H}\left(y_{1:T},p_M(dy_{1:T})\right)$.
	This batch approach would allow approximations using standard Markov chain Monte Carlo
	methods. 
	However, the batch approach is generally not consistent for model selection \citep[see Section 8.1 in][]{dawid2015}.
	Therefore, the prequential
	framework not only has a natural interpretation that relates to sequential probability
	forecasts \citep{dawid1984}, but is also necessary for consistency. This leads to the task of 
	approximating all the successive predictive distributions
	$p(dy_t|y_{1:t-1})$, as described in Section \ref{subsec:smc}. This distinction does not arise for the log-score, for which we always have $-\log p(y_{1:T})=-\sum_{t=1}^T \log p(y_t|y_{1:t-1})$. 
	One consequence of the sequential approach is that different orderings of the observations lead to different sequences of predictive distributions, 
	hence yielding different values of the H-score. This might be undesirable in settings where the observations are not naturally ordered (e.g.\! i.i.d.\! or spatial data). For large samples, this issue is mitigated by the convergence of rescaled H-scores 
	to limits that do not depend on the ordering of the observations (cf. Theorem \ref{theorem:consistencyIID_all_in_one}). For small samples, one could average the H-score over different permutations of the data, or use a random ordering of the data within each SMC run (see Section \ref{sec:consistencyNumericalExample}), at the cost of extra computations.

	\subsection{Numerical illustration with Normal models}
	\label{sec:consistencyNumericalExample}
	
	Inspired by Section 3.2. of \citet{ohagan1995}, we consider the two Normal models
	\begin{align*}
	M_1:&\quad Y_1,...,Y_T\,|\,\theta_1 \stackrel{\text{i.i.d.}}{\sim} \mathcal{N}\left(\theta_1,1^{\vphantom{2}}\right) ,\quad  \theta_1 \sim \mathcal{N}\left(0,\sigma_0^2\right),
	\\
	M_2:&\quad Y_1,...,Y_T\,|\,\theta_2 \,\stackrel{\text{i.i.d.}}{\sim} \mathcal{N}\left(0,\theta_2\right) ,\quad  \theta_2 \sim \text{Inv-}\chi^2\left(\nu_0,s_0^2\right).
	\end{align*}
	The positive hyperparameters are chosen as $\sigma_0^2 =
	10$, $\nu_0 = 0.1$, and $s_0^2 = 1$. We compare $M_1$ and $M_2$, using data generated as $Y_1,...,Y_T \stackrel{\text{i.i.d.}}{\sim} \mathcal{N}(\mu_\star,\sigma_\star^2)$, in the following four settings: (1) $(\mu_\star,\sigma_\star^2)=(1,1)$, i.e.\! $M_1$ is well-specified while $M_2$ is not; (2) $(\mu_\star,\sigma_\star^2)=(0,5)$, i.e.\! $M_2$ is well-specified while $M_1$ is not; (3) $(\mu_\star,\sigma_\star^2)=(4,3)$, i.e.\! both $M_1$ and $M_2$ are misspecified; (4) $(\mu_\star,\sigma_\star^2)=(0,1)$, i.e.\! both $M_1$ and $M_2$ are well-specified.
	
\begin{figure}[h]
		\includegraphics[width=0.85\linewidth]{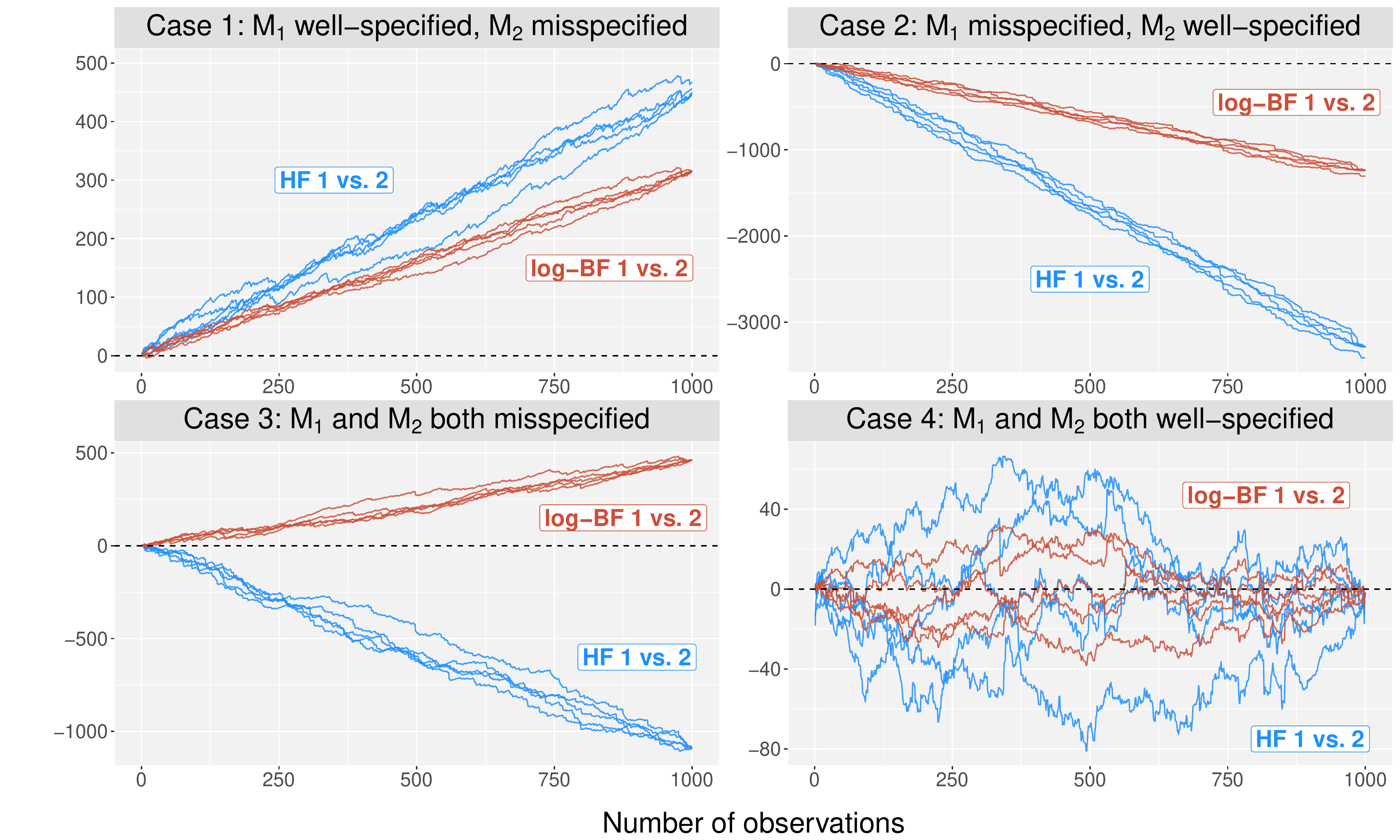}
		\centering
		\caption{Estimated log-Bayes factors (log-BF) and H-factors (HF) of $M_1$ against $M_2$, computed for 5 replications (thin solid lines), under four i.i.d.\! data-generating processes: $\mathcal{N}(1,1)$ (Case 1), $\mathcal{N}(0,5)$ (Case 2), $\mathcal{N}(4,3)$ (Case 3), and $\mathcal{N}(0,1)$ (Case 4). In each plot, the observations are fixed but randomly ordered, so that the variability within each factor is due to Monte Carlo error and random permutation of the data. See Section \ref{sec:consistencyNumericalExample}.}
		\label{fig:iidNormal}
	\end{figure}

\begin{figure}[!h]
		\includegraphics[width=0.83\linewidth]{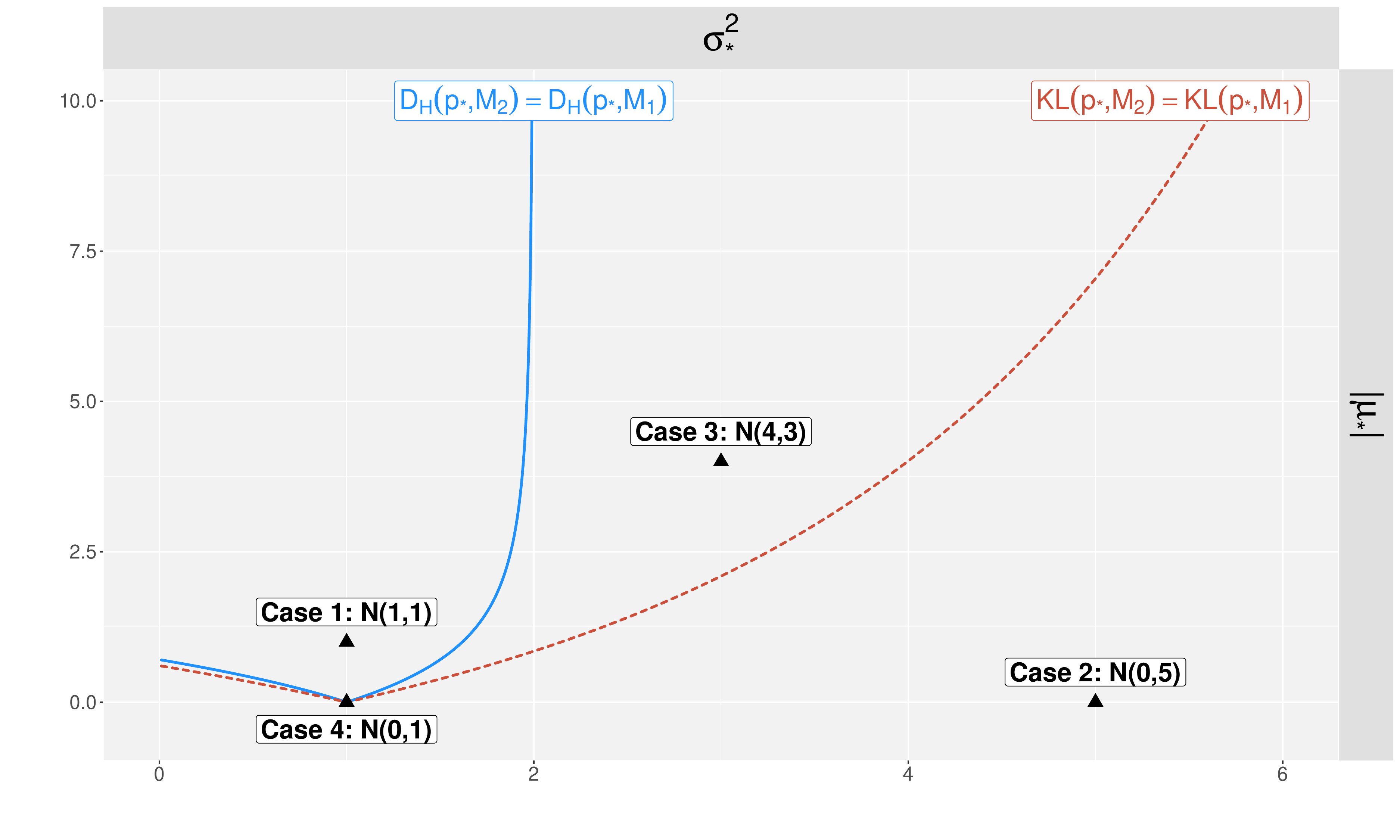}
		\centering
		\caption{Phase plane of $\,d(p_\star,M_2) - d(p_\star,M_1)$ as a function of $(|\mu_\star|,\sigma_\star^2)$, where $d\in\{D_\mathcal{H},\emph{\text{KL}}\}$. The four cases from Section \ref{sec:consistencyNumericalExample} are indicated as triangles. The lines (solid for $D_\mathcal{H}$, dashed for $\emph{\text{KL}}$) are the sets of $(|\mu_\star|,\sigma_\star^2)$ such that $\,d(p_\star,M_2) = d(p_\star,M_1)$. The regions above (resp.\! below) the lines satisfy $\,d(p_\star,M_2) > d(p_\star,M_1)$ (resp.\! $<$), i.e. $M_1$ (resp.\! $M_2$) is closer to $p_\star$.}
		\label{fig:iidNormal_boundaries}
	\end{figure}

	Conjugacy allows all the posterior distributions, scores, and
	divergences to be computed in closed form. The posteriors under
	$M_1$ and $M_2$ concentrate respectively around $\theta^\star_1=\mu_\star$ and
	$\theta^\star_2=\sigma_\star^2 + \mu_\star^2$. We compute $D_\mathcal{H}$ and the
	Kullback-Leibler divergence for Normal densities analytically \citep[see
	Section 6.1 of][]{dawid2015} and get the theoretical limits
	\setlength{\jot}{10pt}
	\begin{align}
	D_\mathcal{H}(p_\star,M_2)-D_\mathcal{H}(p_\star,M_1)&=\frac{\mu_\star^2}{\sigma_\star^2\left(\mu_\star^2+\sigma_\star^2\right)} - \frac{\left(\sigma_\star^2-1\right)^2}{\sigma_\star^2},
	\label{eq:slopeH}
	\\
	\text{KL}(p_\star,M_2)-\text{KL}(p_\star,M_1)&=\frac{1}{2}\log\left(\frac{\mu_\star^2+\sigma_\star^2}{\sigma_\star^2}\right) - \frac{\left(\sigma_\star^2-1\right)-\log\left(\sigma_\star^2\right)}{2},
	\label{eq:slopeLBF}
	\end{align}
	\setlength{\jot}{5pt}
	which depend on the values of $|\mu_\star|$ and $\sigma_\star^2$.
	For each of the four cases, we generate $T=1000$
	observations and perform 5 runs of SMC with $N_\theta=1024$ particles to
	estimate the log-Bayes factors and H-factors of $M_1$ against $M_2$. Each run uses a different ordering of the data, sampled uniformly from all the possible permutations. The
	results are shown in Figure \ref{fig:iidNormal}. H-factors and log-Bayes factors are overlaid on the same plots in order to track their evolution jointly, but their values should not be directly compared.
	As expected in cases 1 and 2,
	the H-factor selects the well-specified model and diverges to infinity at a
	linear rate, with respective slopes matching the theoretical limits 0.5 and
	-3.2 from \eqref{eq:slopeH}. Similar behavior is obtained for the
	log-Bayes factor, which correctly diverges to infinity at the same linear rate,
	with theoretical slopes given by \eqref{eq:slopeLBF}. In case 3, both
	models are misspecified, and \eqref{eq:slopeH}-\eqref{eq:slopeLBF} with
	$(\mu_\star,\sigma_\star^2)=(4,3)$ yield
	$D_\mathcal{H}(p_\star,M_2)-D_\mathcal{H}(p_\star,M_1)\approx -1.05<0$ and
	$\text{KL}(p_\star,M_2)-\text{KL}(p_\star,M_1)\approx 0.47>0$. This leads the
	Bayes factor and the H-factor to favor different misspecified models. 
	In fact, when both $M_1$ and $M_2$ are misspecified, there are infinitely many combinations of $(|\mu_\star|,\sigma_\star^2)\in\mathbb{R}_+^2$ for which $D_\mathcal{H}(p_\star,M_2)<D_\mathcal{H}(p_\star,M_1)$ whereas $\text{KL}(p_\star,M_2)>\text{KL}(p_\star,M_1)$. 
	Indeed, if we define the boundary  $\mathcal{B}_\mathcal{H}(\sigma_\star^2) = |\sigma_\star^2-1|({2-\sigma_\star^2})^{-1/2}$ for $\sigma_\star^2\in(0,2)$ and $\mathcal{B}_\mathcal{H}(\sigma_\star^2) = +\infty$ for $\sigma_\star^2\geq 2$, then $D_\mathcal{H}(p_\star,M_2)=D_\mathcal{H}(p_\star,M_1)$ (resp.\! $>$ and $<$) for $|\mu_\star|=\mathcal{B}_\mathcal{H}(\sigma_\star^2)$ (resp.\! $>$ and $<$). By contrast,  $\text{KL}(p_\star,M_2)=\text{KL}(p_\star,M_1)$ if and only if $|\mu_\star|=\mathcal{B}_\text{KL}(\sigma_\star^2)$, where  $\mathcal{B}_\text{KL}(\sigma_\star^2) = ({\exp\left(\sigma_\star^2-1\right)-\sigma_\star^2})^{1/2}$ for all $\sigma_\star^2>0$. Thus, whenever $\mathcal{B}_\text{KL}(\sigma_\star^2)<|\mu_\star|<\mathcal{B}_\mathcal{H}(\sigma_\star^2)$, the divergences $D_\mathcal{H}$ and $\text{KL}$ disagree on which model is closer to $p_\star$. 
	This is illustrated in Figure \ref{fig:iidNormal_boundaries}. 
	When both divergences are sensible, deciding which one to use would require further considerations \citep[e.g.\! see][]{jewson2018principled}. As explained in Section \ref{sec:intro}, the log-Bayes factor might be inappropriate in the presence of vague priors. Looking back at case 1 for example, since $\log p_{M_1}(y_{1:T})\to -\infty$ when
	$\sigma_0 \to +\infty$, one could always specify a $\sigma_0$ large enough such that the log-Bayes factor would wrongly pick $M_2$. On the other hand, the choice of $M_1$ by the H-factor remains unchanged when $\sigma_0$ increases. This robustness is further illustrated in Section\supphscoreiidNormalrobust of the supplement.
	
	Finally, in case 4, the theoretical slopes are exactly 0, while the models are
	of equal dimensions, hence no model prevails.
	\section{H-score for state-space models}
	\label{sec:hscore_SSM}
	The H-score raises additional computational challenges in the case of state-space models.
	State-space models, also known as hidden Markov models, are a flexible and widely used class of
	time series models \citep{cappe2005,douc:moulines:stoffer2014}, which describe the
	observations $(Y_t)_{t\in\mathbb{N}^*}$ as conditionally
	independent given a latent Markov chain $(X_t)_{t\in\mathbb{N}^*}$ living in $\mathbb{X}\subseteq\mathbb{R}^{d_x}$. A 
	state-space model with parameter
	$\theta\in\mathbb{T}\subseteq\mathbb{R}^{d_\theta}$ specifies an initial
	distribution $\mu_\theta(dx_1)$ of the first state $X_1$, a Markov kernel
	$f_\theta(dx_{t+1}|x_{t})$ for the transition of the latent process, a measurement distribution $g_\theta(dy_t|x_t)$, and a prior distribution
	$p(d\theta)$ on  the parameter. 
	\subsection{Computation of the H-score using SMC$^2$}
	\label{subsec:computation_SMC2}
	The conditional predictive distributions $p(y_t|y_{1:t-1},\theta)$ appearing in
	\eqref{eq:hscoreWithExpectationsMultivariate} correspond to 
	integrals over the latent states, i.e.
	$p(y_t|y_{1:t-1},\theta)=\int
	p(x_t|y_{1:t-1},\theta)\,g_{\theta}(y_t|x_t)\,dx_t$, which are in general intractable.
	Interchanging differentiation and integration under suitable regularity
	conditions yields the following results, which are similar 
	to Fisher's and Louis' identities \citep[Proposition
	10.1.6 in][]{cappe2005}, except 
	that differentiation here is with respect to the observation instead of the parameter. 
	We obtain for all $\theta\in\mathbb{T}$, all observed $y_{1:T}\in\mathbb{Y}^{\,T}$, all $k\in\{1,...,d_y\}$, and all $t\in\{1,...,T\}$,
	{\small
		\begin{align}
		\label{eq:fisher}
		\frac{\partial \log p(y_t|y_{1:t-1},\theta)}{\partial {y_t}_{(k)}} &= \mathbb{E}_t \left[\left.\frac{\partial \log g_\theta(y_t|X_t)}{\partial {y_t}_{(k)}}\right|\theta\right],
		\end{align}}
	\vspace*{-0.5cm}
	\begin{align}
	\resizebox{0.925\textwidth}{!} 
	{
		$\displaystyle\frac{\partial^2 \log p(y_t|y_{1:t-1},\theta)}{\partial {y_t}_{(k)}^{2^{\vphantom{S}}}} + \left(\frac{\partial \log p(y_t|y_{1:t-1},\theta)}{\partial {y_t}_{(k)}}\right)^2
		=\mathbb{E}_t \left[\left.\frac{\partial^2 \log g_\theta(y_t|X_t)}{\partial {y_t}_{(k)}^{2^{\vphantom{S}}}} + \left(\frac{\partial \log g_\theta(y_t|X_t)}{\partial {y_t}_{(k)}}\right)^{\!2\,}\right|\theta\right],
		$
	}
	\label{eq:louis}
	\end{align}
	where the conditional expectations $\mathbb{E}_t$ are with respect to $X_t\sim p(dx_t|y_{1:t},\theta)$.
	Proofs of \eqref{eq:fisher} and \eqref{eq:louis} under regularity assumptions are presented
	in the supplement. Applying \eqref{eq:fisher} and \eqref{eq:louis} to each term in \eqref{eq:hscoreWithExpectationsMultivariate} and using the
	tower property of conditional expectations yields
	\begin{align}
	\resizebox{0.925\textwidth}{!} 
	{
		$\displaystyle\mathcal{H}_{T}(M)=\sum_{t=1}^T\sum_{k=1}^{d_y} \left(2\,\mathbb{E}_t\!\left[\frac{\partial^2 \log g_{\Theta}(y_t|X_t)}{\partial{y_t}_{(k)}^{2^{\vphantom{S}}}} + \left(\frac{\partial \log g_{\Theta}(y_t|X_t)}{\partial {y_t}_{(k)}}\right)^{\!2\,}\right] -\left(\mathbb{E}_t\!\left[\frac{\partial \log g_{\Theta}(y_t|X_t)}{\partial {y_t}_{(k)}}\right]\right)^{\!2\,}\right),
		$
	}
	\label{eq:hscoreExpectationsMultivariateSSM}
	\end{align}
	where the expectations $\mathbb{E}_t$ are with respect to the joint posterior distributions of $(\Theta,X_t)$ given the observations $y_{1:t}$, whose densities are given by $p(\theta,x_t|y_{1:t}) = p(\theta|y_{1:t})p(x_t|y_{1:t},\theta)$.
	
	For many state-space models, the log-derivatives of the measurement density $g_\theta(y|x)$ can be evaluated at any point
	$(\theta,y,x)\in\mathbb{T}\times\mathbb{Y}\times\mathbb{X}$. Assuming that we can simulate 
	the transition kernel of the latent process, we
	can use SMC$^2$ \citep{fulop2013efficient,chopin:jacob:papaspiliopoulos2013} to
	consistently estimate all the conditional expectations appearing in 
	\eqref{eq:hscoreExpectationsMultivariateSSM}. At each time $t\in\{1,...,T\}$,
	SMC$^2$
	produces a set of weighted particles targeting the joint density $p(\theta,x_t|y_{1:t})$, which can be used
	to update the H-score.
	\subsection{Consistency of the H-score for state-space models}
	\label{sec:consistencySSM}
	We revisit the asymptotic consistency results of the H-score in the case of state-space models.
	The observations are no longer assumed to be i.i.d.\! and we
	consider two candidate models, $M_1$ and $M_2$. An additional difficulty in proving 
	consistency of the H-score with dependent observations lies in the approximation of $\mathcal{H}_{T}(M_j)$ by a stationary analog, to which ergodic theorems will apply. As in the i.i.d.\! setting, we
	only give results for univariate continuous observations.

	\begin{theorem}
		\label{theorem:consistencySSM_all_in_one}
		Assume $(Y_t)_{t\in\mathbb{N}^*}$ is ergodic and strongly
		stationary, so that we can artificially extend its set of indices to negative integers and consider the two-sided process $(Y_t)_{t\in\mathbb{Z}}$. Assume $M_1$ and $M_2$ both satisfy the following conditions, where
		models are omitted from the notation and probabilistic statements are $\mathbb{P}_\star$-almost sure:
		\begin{enumerate}[label=(\alph*)]
			\itemsep0.25em 
			\item\label{cond:2a} For all $t\in\mathbb{N}^*$ and $y_{1:t}\in\mathbb{Y}^{t}$, $\theta\mapsto p(y_t|\theta) \, p(\theta|y_{1:t-1})$ is integrable on $\mathbb{T}$. 
			\item\label{cond:2b} For all $t\in\mathbb{N}^*$ and $\theta\in\mathbb{T}$, $y_t\mapsto p(y_t|\theta)$ is twice differentiable on $\mathbb{Y}$.
			\item\label{cond:2c} For all $t\in\mathbb{N}^*$, there exist integrable functions $h_{1,t}$ and $h_{2,t}$ such that, for all $(y_{1:t},\theta)\in\mathbb{Y}^{t}\times\mathbb{T}$, $\left|p(\theta|y_{1:t-1})\,\partial p(y_t|\theta) / \partial {y_t}\right|\leq h_{1,t}(\theta)$ and $\left|p(\theta|y_{1:t-1})\,\partial^2 p(y_t|\theta) / \partial {y_t}^2\right|\leq h_{2,t}(\theta)$.
			\item\label{cond:2d} For all $t\in\mathbb{N}^*$ and $(y_{1:t},\theta)\in\mathbb{Y}^{t}\times\mathbb{T}$, $x_t\mapsto p(x_t|y_{1:t-1},\theta)\,g_\theta(y_t|x_t)$ is integrable on $\mathbb{X}$.
			\item\label{cond:2e} For all $t\in\mathbb{N}^*$ and $(\theta,x_t)\in\mathbb{T}\times\mathbb{X}$, $y_t\mapsto g_\theta(y_t|x_t)$ is twice differentiable on $\mathbb{Y}$.
			\item\label{cond:2f} There exist integrable functions $h_{3,t}$ and $h_{4,t}$ such that, for all $(y_{1:t},\theta,x_t)\in\mathbb{Y}^{t}\times\mathbb{T}\times\mathbb{X}$, $\left|p(x_t|y_{1:t-1},\theta) {\partial g_\theta(y_t|x_t)}/{\partial {y_t}}\right|\leq h_{3,t}(x_t)$ and $\left|p(x_t|y_{1:t-1},\theta){\partial^{2} g_\theta(y_t|x_t)}/{\partial {y_t^2}} \right|\leq h_{4,t}(x_t)$.
			\item\label{cond:2g} For all $t\in\mathbb{N}^*$, there exists $\theta^\star\in\mathbb{T}$ such that, if $\Theta_t\sim p(d\theta|Y_{1:t})$ for all $t\in\mathbb{N}^*$, then $\Theta_t\xxrightarrow[t \to +\infty]{\mathcal{D}}\theta^\star$.
			\item\label{cond:2h} There exist a constant $L>0$ and a neighborhood $\,\mathcal{U}_{\theta^\star}$ of $\theta^\star$ such that, for all $t\in\mathbb{N}^*$, $\theta\mapsto\mathcal{H}\left(Y_t,p(dy_t|Y_{1:t-1},\theta)\right)$  and $\theta\mapsto \partial \log p(Y_t|Y_{1:t-1},\theta)/\partial y_t$ are $L$-Lipschitz functions.
			\item\label{cond:2i} There exist $\alpha_1>1$ and $\alpha_2>1$ such that $\,\sup_{t\in\mathbb{N}^*}\mathbb{E}\left[\,|\mathcal{H}\left(Y_t,p(dy_t|Y_{1:t-1},\Theta_t)\right)|^{\alpha_1}\,|\,Y_{1:t}\,\right]< +\infty$ and $\,\sup_{t\in\mathbb{N}^*}\mathbb{E}\left[\left(\partial \log p(Y_t|Y_{1:t-1},\Theta_t)/\partial y_t\right)^{2\,\alpha_2}\,|\,Y_{1:t}\,\right]< +\infty$, where the conditional expectations are with respect to the posterior distribution $\Theta_t\sim p(d\theta|Y_{1:t})$.
			\item\label{cond:2j} There is a dominating probability measure $\eta$ on $\mathbb{X}$ such that the transition kernel $f_{\theta^\star}(dx_{t+1}|x_{t})$ has density $\nu_{\theta^\star}(x_{t+1}|x_t)=(df_{\theta^\star}(\cdot|x_{t})/d\eta)(x_{t+1})$ with respect to $\eta$.
			\item\label{cond:2k} There exist positive constants $\sigma^-$ and $\sigma^+$ such that, for all $(x_t,x_{t+1})\in\mathbb{X}\times\mathbb{X}$, the transition density $\nu_{\theta^\star}(x_{t+1}|x_t)$ satisfies $0<\sigma^-<\nu_{\theta^\star}(x_{t+1}|x_t)<\sigma^+<+\infty$.
			\item\label{cond:2l} For all $y_t\in\mathbb{Y}$, the integral $\int_\mathbb{X}g_{\theta^\star}(y_t,x_t)\,\eta(dx_t)$ is bounded away from $0$ and $+\infty$.
			\item\label{cond:2m} ${b_{}}= \sup_{\substack{x\in\mathbb{X}\\y\in\mathbb{Y}}}\limits \left|\frac{\partial^2\log g_{\theta^\star}(y|x)}{\partial y^{2^{\vphantom{S}}}} + \left(\frac{\partial\log g_{\theta^\star}(y|x)}{\partial y^{\vphantom{\mathcal{S}^S}}}\right)^2\right|<+\infty\,$ and $\,{c_{}} = \sup_{\substack{x\in\mathbb{X}\\y\in\mathbb{Y}}}\limits \left|\frac{\partial\log g_{\theta^\star}(y|x)}{\partial y^{\vphantom{\mathcal{S}^S}}}\right| < +\infty$.
			\item\label{cond:2n} $\sup_{\substack{x\in\mathbb{X}\\y\in\mathbb{Y}}}\limits g_{\theta^\star}(y|x)<+\infty\,$ and $\,\mathbb{E}_\star\left[\left|\log\left(\int_\mathbb{X}g_{\theta^\star}(Y_1|x)\nu_{\theta^\star}(dx)\right)\right|\right] <+\infty$.
			\item\label{cond:2o} The conditional density $y_1\mapsto p_\star(y_1|Y_{-\infty:0})$ of $Y_1$ given $(Y_t)_{t\leq 0}$ is well-defined and twice differentiable, and $\mathbb{E}_\star\left[\left|\mathcal{H}\left(Y_1,p_\star(dy_1|Y_{-\infty:0})\right)\right|^{\vphantom{S}}\right]< +\infty$.
		\end{enumerate} 
		If these conditions are met, we may define, for each $j\in\{1,2\}$, the quantity
		\begin{align}
		\label{eq:DivergenceDeltaDefinitionSSM}
		D_\mathcal{H}(p_\star,M_j)=\mathbb{E}_\star\left[\mathcal{H}\left(Y_1,p_{j}(dy_1|Y_{-\infty:0},\theta^\star_j)\right)^{\vphantom{S}}\right] - \mathbb{E}_\star\left[\mathcal{H}\left(Y_1,p_\star(dy_1|Y_{-\infty:0})\right)^{\vphantom{S}}\right],
		\end{align}
		where $p_{j}(y_1|Y_{-\infty:0},\theta^\star_j)$ is the provably well-defined conditional density of $Y_1$ given $(Y_t)_{t\leq 0}$ under $M_j$ and $\theta^\star_j$. Under these conditions, we have
		\begin{align}
		\label{theorem:consistencySSM}
		\frac{1}{T}\left(\mathcal{H}_{T}({M_2})-\mathcal{H}_{T}({M_1})^{\vphantom{S^S}}\right) \;\xxrightarrow[T \to +\infty]{\mathbb{P}_\star-a.s.}\;D_\mathcal{H}(p_\star,M_2)-D_\mathcal{H}(p_\star,M_1)\,.
		\end{align}
		If $\,p_\star(y_1|Y_{-\infty:0}) \,{\partial \log p(y_1|Y_{-\infty:0},\theta^\star)}/{\partial y_1} \xxrightarrow[|y_1|\to +\infty]{\mathbb{P}_\star-a.s.} 0$, then $D_\mathcal{H}(p_\star,M_j)\geq 0$, with $D_\mathcal{H}(p_\star,M_j)=0$ if and only if $p_{j}(y_1|Y_{-\infty:0},\theta^\star_j)=p_\star(y_1|Y_{-\infty:0})$, $\mathbb{P}_\star$-almost surely. 
	\end{theorem} 
	Conditions \ref{cond:2a} to \ref{cond:2c} ensure the validity of \eqref{eq:prequentialHscoreExpectationVariance}; \ref{cond:2d} to \ref{cond:2f} ensure the validity of \eqref{eq:fisher} and \eqref{eq:louis}; \ref{cond:2g} assumes the concentration of the posterior to a point mass; \ref{cond:2h} to \ref{cond:2i} yield suitable convergence of posterior moments; \ref{cond:2j} to \ref{cond:2l} ensure the \textit{forgetting propriety} of the latent Markov chain and the H-score; \ref{cond:2m} to \ref{cond:2n} relate to the well-definiteness of the conditional density $p_{j}(y_1|Y_{-\infty:0},\theta^\star_j)$; finally, \ref{cond:2o} and the last boundary condition ensure that the H-score is strictly proper and well-defined for $p_\star$. Further discussion on these conditions and detailed proofs are provided in Section\suppconsistency of the supplement. 
	
	For state-space models, posterior concentration results have been derived in specific
	cases \citep[e.g.][and references
	therein]{lijoi2007bayesian,de2008asymptotic,shalizi2009dynamics,gassiat2014posterior,douc:moulines:stoffer2014,douc:olsson:roueff}.
	However, to the best of our knowledge, general results on posterior
	concentration for misspecified state-space models have yet to be
	established. As a consequence, our proof of Theorem \ref{theorem:consistencySSM_all_in_one} uses posterior concentration as a working
	assumption. Our numerical examples
	suggest that concentration of posterior distributions can be observed in practice, even for complex state-space models (see posterior density plots in Section\suppdensityplots of the supplement). 
	Further research on Bayesian asymptotics in state-space models might provide more theoretical understanding of such phenomena.
	
	\subsection{Illustration with L\'{e}vy-driven stochastic volatility models}
	\label{example:applicationsSV}
	In this simulation study we illustrate the consistency of the H-score
	in nonlinear, non-Gaussian state-space models with continuous
	observations. 
	A simpler example with linear Gaussian state-space and ARMA models can be found
	in Section\suppconsistencyTimeSeries of the supplement. Here we
	consider L\'{e}vy-driven stochastic volatility models 
	\citep{barndorff2011,bns:real}. These models feature intractable transition kernels 
	that can only be simulated, and describe the joint
	evolution of
	the log-returns $Y_t$ and the instantaneous volatility $V_t$ of a financial
	asset. The former is modeled as a continuous time process driven by a Brownian
	motion, while the latter is modeled as a L\'{e}vy process. 
	Given a triplet $(\lambda,\xi,\omega)$, we can generate random
	variables $(V_{t},Z_{t})_{t\geq 1}$ recursively as:
	{
		\begin{align}
		\left.
		\begin{array}{ll}
		\!k \sim \text{Poisson}\left(\lambda\xi^2/\omega^2\right) ;\quad  C_{1:k}\,\stackrel{\text{i.i.d.}}{\sim}\,\text{Unif}(t-1,t) ;\quad  E_{1:k}\,\stackrel{\text{i.i.d.}}{\sim}\,\text{Exp}\left(\xi/\omega^2\right) ;
		\\
		\!Z_0 \sim\text{Gamma}\left({\xi^2}/{\omega^2},\,{\xi}/{\omega^2}\right) ;\quad Z_{t} = e^{-\lambda}Z_{t-1} + \sum_{j=1}^k e^{-\lambda\left(t-C_j\right)}E_j ;
		\\  
		V_{t} = \frac{1}{\lambda}\left(Z_{t-1}-Z_{t}+\sum_{j=1}^k E_j\right).
		\end{array}
		\right\}
		\label{eq:SVstateEquation}
		\end{align}
	}
	The first model ($M_1$) describes the volatility as driven by a \emph{single factor}, expressed in terms of a finite rate Poisson process.
	\begin{itemize}
		\item [$M_1$:]\textit{$(V_{t},Z_{t})$ from \eqref{eq:SVstateEquation} given $(\lambda,\,\xi,\,\omega)$; \quad $X_{t}=(V_{t},Z_{t})$;\quad $Y_{t}\,|\,X_t \sim \mathcal{N}\left(\mu + \beta V_{t}, V_{t}\right)$\,; \quad with independent priors \,$\lambda\sim\emph{\text{Exp}}(1);\quad \xi,\,\omega^2 \stackrel{\emph{\text{i.i.d.}}}{\sim} \emph{\text{Exp}}\left(1/5\right) ;\quad \mu,\,\beta \stackrel{\emph{\text{i.i.d.}}}{\sim} \mathcal{N}(0,10)$.}
	\end{itemize}
	The second model ($M_2$) introduces an additional independent component to drive the behavior of the volatility, leading to the \emph{multi-factor} model below.
	\begin{itemize}
		\item [$M_2$:]\textit{$(V_{i,t}\,,Z_{i,t})$ from \eqref{eq:SVstateEquation} independently for $i\in\{1,2\}$ given $\left(\lambda_i,\xi\text{\emph{w}}_i,\omega\text{\emph{w}}_i\right)$, with $(\emph{\text{w}}_1,\emph{\text{w}}_2) = (\emph{\text{w}},1-\emph{\text{w}})$;\quad $X_{t}=(V_{1,t}\,,V_{2,t}\,,Z_{1,t}\,,Z_{2,t})$;\quad $Y_{t}\,|\,X_t \sim \mathcal{N}\left(\mu + \beta V_{t}, V_{t}\right)$ where $V_{t} = V_{1,t} + V_{2,t}$\,;\quad with independent priors \, $\lambda_1\sim\emph{\text{Exp}}(1)$;\quad $\lambda_2-\lambda_1\sim\emph{\text{Exp}}\left(1/2\right)$;\quad $\emph{\text{w}}\sim\emph{\text{Unif}}(0,1)$;\quad $\xi,\,\omega^2 \stackrel{\emph{\text{i.i.d.}}}{\sim}\emph{\text{Exp}}\left(1/5\right)$ ;\quad $\mu,\,\beta \stackrel{\emph{\text{i.i.d.}}}{\sim} \mathcal{N}(0,10)$.}
	\end{itemize}
	
	For model $M_1$, we can prove that there exist values of the parameter $\theta=(\lambda,\xi,\omega,\mu,\beta)$ such that $\mathbb{E}[\,|\partial \log
	g_\theta (y_1|X_1) / \partial y_1|\,] = +\infty$, which prevents the use of
	\eqref{eq:fisher}-\eqref{eq:louis} to estimate the H-score of model $M_1$.
	When \eqref{eq:fisher}-\eqref{eq:louis} do not hold, 
	we can directly estimate the partial derivatives of $\tilde{y}_t\mapsto p(\tilde{y}_t|y_{1:t-1},\theta)$
	at the observed $y_t$, by using approximate draws from the conditional predictive distribution $p(dy_t|y_{1:t-1},\theta)$. Approximate draws from $p(dy_t|y_{1:t-1},\theta)$
	can be obtained from a run of SMC\textsuperscript{\,2}, as long as one can sample from the measurement distribution $g_\theta(dy_t|x_t)$. For a chosen bandwidth $h>0$ (e.g. \citeauthor{hardle1990bandwidth}, \citeyear{hardle1990bandwidth}, and references in Section 1.11 of \citeauthor{tsybakov2009introduction}, \citeyear{tsybakov2009introduction}) and a twice continuously differentiable kernel $K$
	integrating to 1, e.g.\! a standard Gaussian kernel
	$K(u)=(2\pi)^{-1/2}\exp(-u^2/2)$, we can use $n$ draws
	$\tilde{y}_{t}^{(1)},\dots,\tilde{y}_{t}^{(n)}$ from $p(dy_t|y_{1:t-1},\theta)$
	to consistently estimate $p(y_t|y_{1:t-1},\theta)$ by the kernel density
	estimator $\widehat{p}(y_t|y_{1:t-1},\theta)=(nh)^{-1}\sum_{i=1}^n
	K\!(({y_t-\tilde{y}_t^{(i)}})/{h})$.
	This kernel density estimator is twice differentiable with respect to $y_t$, hence we can respectively use ${\partial\,\widehat{p}({y}_t|y_{1:t-1},\theta)}/{\partial {{y_t}}_{(k)}}$ and ${\partial^2\,\widehat{p}({y}_t|y_{1:t-1},\theta)}/{\partial {{y_t}}_{(k)}^2}$ as consistent estimators of the partial derivatives ${\partial p({y}_t|y_{1:t-1},\theta)}/{\partial {{y_t}}_{(k)}}$ and ${\partial^2 p({y}_t|y_{1:t-1},\theta)}/{\partial {{y_t}}_{(k)}^2}$, as $n\to+\infty$ and $h\to 0$ at an appropriate rate \citep[e.g.][]{bhattacharya1967estimation}.
	
\begin{figure}[h]
		\includegraphics[width=0.85\linewidth]{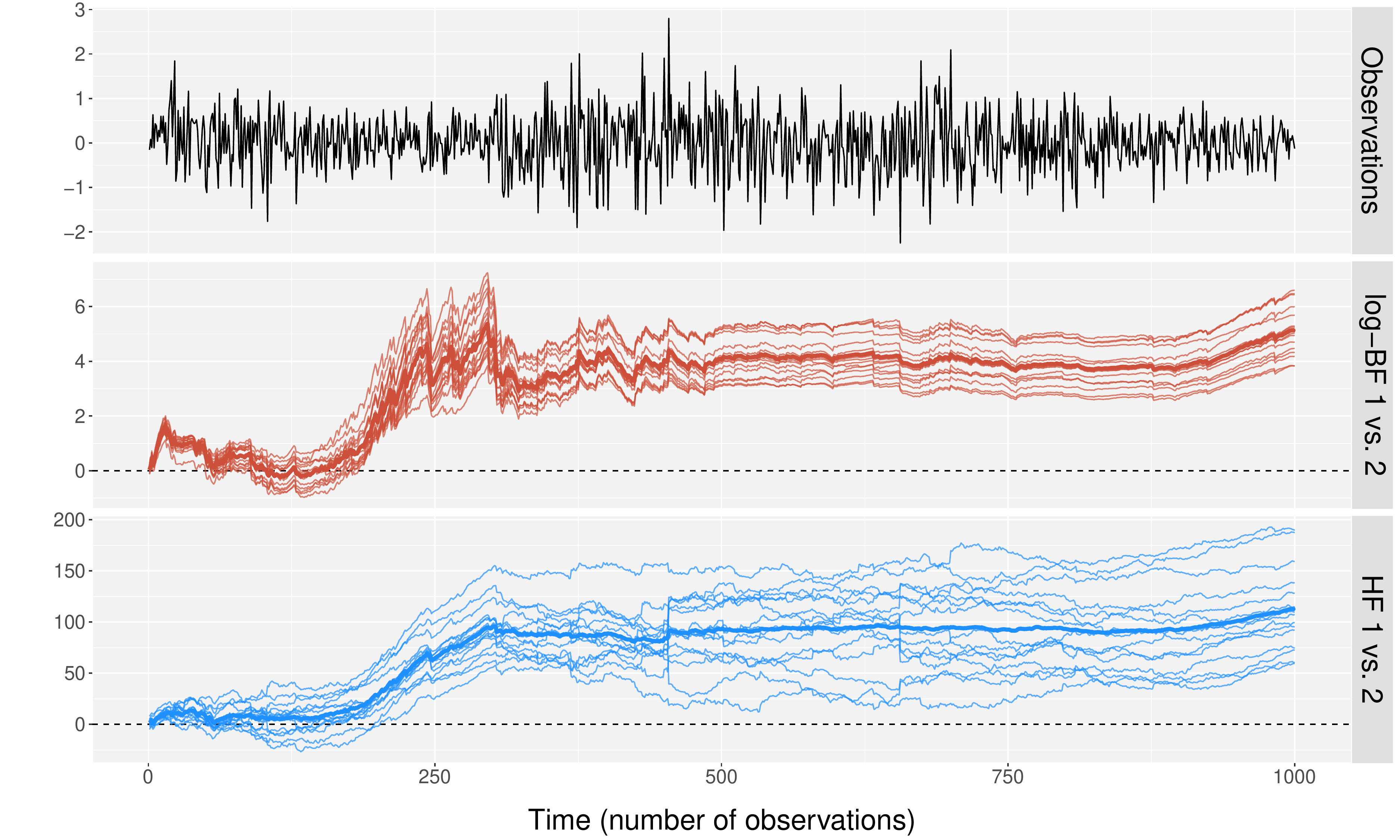}
		\centering
		\caption{Top panel: log-returns simulated from model $M_1$ with parameters $\lambda=0.01$, $\xi = 0.5$, $\omega^2=0.0625$, $\mu=0$, and $\beta=0$. Middle and bottom panels: estimated log-Bayes factor (log-BF) and H-factor (HF) of $M_1$ against $M_2$, computed for 15 replications (thin solid lines), along with the average scores across replications (thick solid lines). In each plot, the variability within each factor is due to Monte Carlo error. See Section \ref{example:applicationsSV}.}
		\label{fig:SV_AllinOne}
	\end{figure}

	We simulate $T=1000$ observations from a single-factor L\'{e}vy-driven
	stochastic volatility model with parameters $\lambda=0.01$, $\xi = 0.5$,
	$\omega^2=0.0625$, $\mu=0$, and $\beta=0$, following the simulations of  \citet{bns:real}. 
	The H-factor of $M_1$ against
	$M_2$ is computed for 15 replications of
	SMC$^2$, using $N_\theta = 1024$ particles in $\theta$, and an adaptive number
	of particles in $x$ starting at $N_x = 128$. The kernel density estimation is
	performed with a Gaussian kernel, using $n = 1024$ predictive draws and
	$h=0.1$. 
	The estimated log-Bayes factor and H-factor of $M_1$ against $M_2$ are plotted
	in Figure \ref{fig:SV_AllinOne}. Here the models are nested and well-specified, but their dimensions differ.  We see that both criteria
	correctly select the smaller model $M_1$. As mentioned in Section \ref{subsec:smc}, the estimated H-factor tends to have a larger relative variance than the estimated log-Bayes factor, especially in the presence of extreme observations (e.g.\! at times $454$ and $656$), and might thus call for a larger number of particles.
	\section{H-score for discrete observations}
	\label{sec:discrete}
	Motivated by an application in population dynamics (Section
	\ref{example:applicationsKangaroos}), we propose an extension of the H-score to
	discrete observations. We assume that each observation
	$y=(y_{(1)},...,y_{(d_y)})^\top$ takes finite
	values (i.e.\! $\|y\|<+\infty$) in some discrete space $\mathbb{Y}=\llbracket
	a_1,b_1\, \rrbracket\times...\times\llbracket a_{d_y},b_{d_y}\, \rrbracket$,
	where ${\llbracket a_k,b_k\, \rrbracket=[a_k,b_k]}\cap\mathbb{Z}$ and ${a_k,b_k
		\in \mathbb{Z}\cup\{-\infty,+\infty\}}$, with $a_k<b_k$ for all
	$k\in\{1,...,d_y\}$. For ease of exposition, assume for now that $b_k-a_k\geq 3$ for all
	$k\in\{1,...,d_y\}$.
	\subsection{Extension of the H-score to discrete observations}
	\label{subsec:extension_hscore_discrete}
	Let $e_k$ denote the canonical vector of $\mathbb{Z}^{d_y}$ that has all coordinates equal to $0$ except for its
	$k$-th coordinate that equals $1$. For all $y\in\mathbb{Y}$, all non-negative functions $p$ on $\mathbb{Y}$, and all $k\in\{1,...,d_y\}$, we define $\partial_k\, p(y) = (p(y+e_k)-p(y-e_k))/2$ and $\partial_k \log p(y) = \partial_k\, p(y)/p(y)$. We define the score
	\begin{align}
	{\mathcal{H}^D}(y,p)=\sum_{k=1}^{d_y} \mathcal{H}^D_{k}(y,p)\,,
	\end{align}
	where $\mathcal{H}^D_{k}(y,p) = 2\,\partial_k\left(\partial_k \log p(y)\right) + \left(\partial_k \log p(y)^{\vphantom{S}}\right)^2$ if $a_k+2\leq y_{(k)} \leq b_k-2$. At the boundaries, we define $\mathcal{H}^D_{k}(y,p)$ respectively as $\partial_k \log p(y+e_k)$, $\partial_k \log p(y+e_k) + \left(\partial_k \log p(y)\right)^2$, $-\partial_k \log p(y-e_k) + \left(\partial_k \log p(y)\right)^2$, and $-\partial_k \log p(y-e_k)$ for $y\in\{a_k,a_k+1,b_k-1,b_k\}$. 
	
	The expression of ${\mathcal{H}_k^D}$ can be regarded as a discrete analog of
	the H-score where the partial derivatives are replaced by central finite
	differences. Proper scores for discrete observations can be entirely
	characterized as super-gradients of concave entropy functions
	\citep{McCarthy1956, hendrickson1971,dawid2012}. Using this characterization,
	we can prove that ${\mathcal{H}^D}$ is a proper scoring rule.  
	
	If $b_k = a_k + 1$ (e.g. for binary data) or $b_k = a_k + 2$, we could still
	define $\mathcal{H}^D_{k}$  by ignoring the cases
	$y_{(k)}=a_k+1$, or $y_{(k)}=b_k-1$, or both. Alternatively, we could use
	forward differences. All these definitions lead to scores that meet the requirements of
	being insensitive to prior vagueness, while being proper and local. Deciding
	which one to use is then a matter of further considerations, left for future research. The construction of ${\mathcal{H}^D}$ and
	the proof of its propriety are detailed in Section\suppdiscretehscore of the supplement. 
	
	\subsection{Diffusion models for population dynamics of red kangaroos}
	\label{example:applicationsKangaroos}
	
\begin{figure}[h]
		\includegraphics[width=0.85\linewidth]{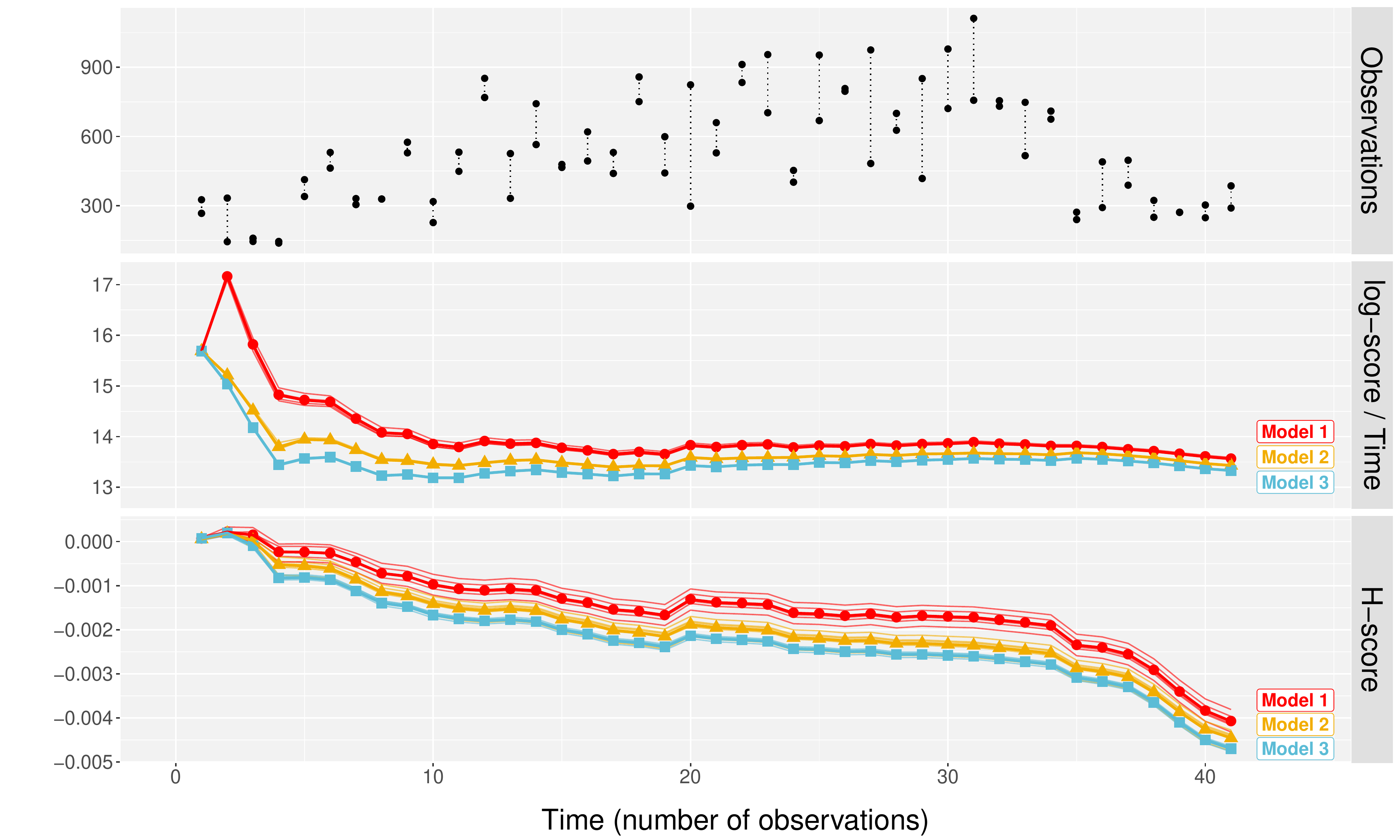}
		\centering
		\caption{Top panel: double transect counts of red kangaroos. Middle and bottom panels: estimated log-scores and H-scores of $M_1$ (circles), $M_2$ (triangles), and $M_3$ (squares), for 5 replications (thin solid lines), along with the average scores across replications (thick lines with shapes). The log-scores are rescaled by the number of observations for better readability. The variability within each model is due to Monte Carlo error. See Section \ref{example:applicationsKangaroos}.}
		\label{fig:kangarooAllinOne}
	\end{figure}

	We illustrate the H-score for discrete observations
	by comparing three nonlinear non-Gaussian state-space models, describing the
	dynamics of a population of red kangaroos (\emph{Macropus rufus}) in New South
	Wales, Australia. These models were compared in \citet{knape2012} using Bayes
	factors, although the authors acknowledged the undesirable sensitivity of their
	results to their choice of prior distributions. The data
	\citep[][]{caughley1981} is a time series of 41 bi-variate observations
	$(Y_{1,t}\,,Y_{2,t})$, formed by double transect counts of red kangaroos,
	measured between
	1973 and 1984 (see Figure \ref{fig:kangarooAllinOne}). 
	The small number of observations calls for a criterion
	that is principled for finite samples, contrarily to e.g. the Bayesian Information Criterion.
	The models are nested
	and will be referred to as $M_1$, $M_2$, and $M_3$, by decreasing order of
	complexity. The largest model ($M_1$) is a logistic diffusion model. Simpler
	versions include an exponential growth model ($M_2$) and a random-walk model
	($M_3$). In these models a latent population size $(X_t)$ follows a
	stochastic differential equation \citep[see motivation in][]{dennis1988,knape2012}. 
	Each model is specified  below, where $(W_t)_{t\geq 0}$ denotes a standard Brownian motion. 
	{\small
		\begin{itemize}
			\item [$M_1$:]\textit{$X_1 \sim \emph{\text{LN}}(0,5) \,;\quad  dX_t/X_t=(\sigma^2/2+r-bX_t)\,dt+\sigma dW_t \,;$
				\\
				$Y_{1,t}\,,Y_{2,t}\,|\,X_t\,,\tau\,\stackrel{\emph{\text{i.i.d.}}}{\sim}\,\emph{\text{NB}}(X_t,\,X_t+\tau X_t^2)\,;$
				\\
				with independent priors; $\sigma,\,\tau,\,b \stackrel{\emph{\text{i.i.d.}}}{\sim} \emph{\text{Unif}}(0,10),\;r\sim\emph{\text{Unif}}(-10,10)$.}
			\item [$M_2$:]\textit{same as $M_1$ with $b=0$; with independent priors $\sigma,\,\tau \stackrel{\emph{\text{i.i.d.}}}{\sim} \emph{\text{Unif}}(0,10),\;r\sim\emph{\text{Unif}}(-10,10)$.}
			\item [$M_3$:]\textit{same as $M_1$ with $b=0$ and $r=0$; with independent priors $\sigma,\,\tau \stackrel{\emph{\text{i.i.d.}}}{\sim} \emph{\text{Unif}}(0,10)$.}
		\end{itemize}
	}
	We perform 5 runs of SMC$^2$
	to estimate the log-score and H-score of
	each model, with an adaptive number $N_x$ of latent particles. We use $N_\theta =16384$ particles in
	$\theta$, and $N_x = 32$ initial particles in $x$.  For model
	$M_1$, we simulate the latent process using the Euler-Maruyama method with
	discretization step $\Delta_t=0.001$. The estimated log-scores and H-scores are shown in Figure
	\ref{fig:kangarooAllinOne}. For better readability, the log-score is rescaled by the number of observations. Using the H-scores would lead to selecting
	model $M_3$, similarly to \citet{knape2012} who use log-scores.  
	Their conclusion was mitigated by the sensitivity of the evidence to the choice of vague priors: for instance,
	changing the prior on $r$ in model $M_2$ to $\text{Unif}(-100,100)$ effectively divides the evidence of $M_2$ by a factor 10.
	On the other hand, we have found the impact of that change of prior on the H-score to be indistinguishable from the Monte Carlo variation
	across runs.
	
	\section{Discussion}
	\label{sec:discussion}
	
	The H-factor constitutes a competitive alternative to the Bayes factor. 
	Not only is it justified non-asymptotically since it relies on assessing predictive performances using a proper local scoring rule, but it is robust to the arbitrary vagueness
	of prior distributions. It can be applied to a large variety of
	models --- including nonlinear non-Gaussian state-space models --- and
	it can be estimated sequentially with SMC or SMC$^2$, at a cost comparable to
	that of the Bayes factor. Using our R implementation, one SMC or SMC$^2$ replication took about a few minutes for each i.i.d.\! Normal models with $1000$ observations, about an hour for each kangaroo population model with $41$ observations, and about five hours for each stochastic volatility model with $1000$ observations. In all cases, the Monte Carlo error can be arbitrarily reduced
	by increasing the number of particles $N_\theta$ \citep[Section 3 in][]{chopin:jacob:papaspiliopoulos2013}. However, the H-score puts
	additional smoothness restrictions on the models, e.g.\! the twice
	differentiability of their predictive distributions with respect to the
	observations \citep[see][and its rejoinder]{dawid2015}. Thus there are 
	models for which the Bayes factor is applicable but not the H-factor. 
	We have also discussed in Section \ref{sec:consistencyNumericalExample} a case
	where the two criteria disagree, even asymptotically, contrarily to
	e.g. partial and intrinsic Bayes factors \citep{santis1999methods}
	that asymptotically agree with the Bayes factor.
	
	The sequential form of the score is problematic when observations are not naturally ordered, leading to different values
	of the H-score for different orderings. This issue is mitigated by the
	following facts: if the sample is large enough, any ordering of the data would yield similar H-scores.
	For smaller samples, one could average the H-score over
	random permutations of the data. In that case, quantifying and controlling the extra
	variability induced by these permutations would deserve investigation. 
	
	For continuous observations and non-nested parametric models satisfying strong
	regularity assumptions, we have proved that the H-score leads to consistent
	model selection. The asymptotic behavior of the H-factor is determined by how
	close the candidate models are from the data-generating process, where 
	closeness is quantified by the relative Fisher information divergence associated with the H-score, in
	contrast to the Kullback–Leibler divergence associated with the Bayes factor. Our proofs rely on strong assumptions, but the numerical experiments indicate 
	that the results might hold in more generality. Results for discrete observations and nested
	well-specified models would be interesting topics of future research. 
	It would be interesting to study frequentist properties of the proposed model choice procedure, 
	e.g.\! by deriving confidence intervals for the difference in expected H-scores. 
	One could for instance complement the results of Theorems
	\ref{theorem:consistencyIID_all_in_one} and
	\ref{theorem:consistencySSM_all_in_one} with central limit
	theorems, which would enable further connections between Bayesian model selection criteria
	and likelihood ratio tests as described e.g.\! in \citet{vuong1989likelihood}.
	
	To deal with vague or improper priors, other alternatives to the log-evidence include Bayesian
	cross-validation criteria, e.g. $\sum_{t=1}^T \log p(y_t|y_{-t})$, where $y_{-t} = \{y_s:1\leq s\leq T
	\text{ and } s\neq t\}$. Such criteria would be
	applicable under weaker smoothness assumptions on the predictive densities,
	while still being robust to arbitrary vagueness of
	prior distributions. Efficient computation of these criteria
	is challenging,
	and can be envisioned for i.i.d.\! models using MCMC \citep{alqallaf2001cross} or SMC methods
	\citep{bornn2010efficient}; the case of state-space models would
	be more challenging, due to standard difficulties arising when splitting time series. Another approach suggested in
	\citet{kamary:mengersen:robert:rousseau2014} is to cast model selection as a
	mixture estimation problem, which also raises questions in the case of time series.
	\bibliographystyle{my_agsm}
	\bibliography{biblio}
	
\end{document}